\documentclass[letterpaper,english,reprint,aps,floatfix]{revtex4-1}
\usepackage{amsmath}
\usepackage{amssymb}
\usepackage{graphicx}
\usepackage{array}
\usepackage{dcolumn}
\usepackage{subfigure}
\usepackage{color}
\usepackage{float}
\usepackage{natbib}
\usepackage{epstopdf}
\usepackage[T1]{fontenc}
\usepackage[utf8]{inputenc}
\usepackage[hidelinks=true]{hyperref}
\usepackage[normalem]{ulem}
\hypersetup{
  colorlinks   = true, % Colours links instead of ugly boxes
  urlcolor     = blue, % Colour for external hyperlinks
  linkcolor    = blue, % Colour of internal links
  citecolor    = red   % Colour of citations
}
\makeatletter

\providecommand{\tabularnewline}{\\}
\makeatother
\usepackage{babel}

\begin{document}
\preprint{APS/123-QED}
\title{Persistent Spin Textures in Nonpolar Chiral Systems}
\author{Kunal Dutta}
\email{pskd2298@iacs.res.in}
\author{Indra Dasgupta}
\email{sspid@iacs.res.in}
\affiliation{School of Physical Sciences, Indian Association for the Cultivation
of Science, 2A and 2B Raja S.C. Mullick Road, Jadavpur, Kolkata 700
032, India}
\date{\today}
%%%%%%%%%%%%%%%%%%%%%%%%%%%%%%%%%%%%%%%%%%%%%%%%%%%%%%%%%%%%%%%%%%%%%%%%%%%%%%
\begin{abstract}

In this {paper}, we have proposed a novel route for the realisation of persistent spin texture (PST). We have shown from symmetry considerations that in non-polar chiral systems, bands with specific orbital characters around a high symmetry point with $D_{2}$ little group may admit a single spin dependent term in the low energy $\bf{k.p}$ model Hamiltonian that naturally leads to PST. Considering a $2D$ plane in the Brillouin zone (BZ), we have further argued  that in such chiral systems the  PST is transpired due to the comparable strengths of the  Dresselhaus and Weyl (radial) interaction parameters where the presence of these two terms are allowed by the $D{_2}$ symmetry. Finally using first principles density functional theory (DFT) calculations we have identified that the non-polar chiral compounds Y$_3$TaO$_7$ and {AsBr$_3$} displays PST for the conduction band and {valence band respectively} around the $\Gamma$ point having $D{_2}$ little group and predominantly  Ta-$d_{xz}$ orbital  {character for Y$_3$TaO$_7$ and Br-$p{_x}$ orbital character for AsBr$_3$}  corroborating our general strategy. Our results for the realisation of PST in non-polar chiral systems thereby broaden the class of materials displaying PST that can be employed for application in spin-orbitronics.
\end{abstract}
\maketitle

%%%%%%%%%%%%%%%%%%%%%%%%%%%%%%%%%%%%%%%%%%%%%%%%%%%%%%%%%%%%%%%%%%%%%%%%%%%%%%
In the recent times, non-centrosymmetric systems have received considerable interest where the spin degeneracy of the bands may be lifted in an otherwise non-magnetic system due to the presence of a momentum dependent magnetic field $\bold{\Omega}\left(\bold{k}\right)$ in the rest frame of the electron. The momentum dependent field $\bold{\Omega}\left(\bold{k}\right)$ further locks the electron's spin direction to its momentum and leads to complex spin textures in the reciprocal space \citep{Manchon2015} primarily enforced by  symmetries \citep{PhysRevB.104.104408}. Depending on the symmetry, the system may display Rashba \citep{rashba1960properties,Bihlmayer_2015}, Dresselhaus \citep{PhysRev.100.580}, Weyl (radial) \citep{PhysRevLett.114.206401}, persistent \citep{PhysRevLett.90.146801, RevModPhys.89.011001} or more complex spin configurations in the momentum space. Of particular interest for application in spin-orbitronics are  spin textures that  maintain a uniform spin-configuration in the momentum space due to an unidirectional spin-orbital  field  $\left(\bf{\Omega}\right)$ referred to as  persistent spin texture (PST). The motion of electrons in such a scenario  is accompanied by spin precession around this unidirectional field, giving rise to a spatially periodic mode known as  persistent spin helix (PSH) \citep{PhysRevLett.97.236601}. This distinctive PSH state emerges due to the presence of SU(2) spin rotation symmetry, which exhibits resilience against spin-independent disorder and  essentially has the potential to provide an infinite spin life-time \citep{PhysRevLett.117.236801}.
\\
\indent A manifestation of PST usually transpires when a system exhibits comparable magnitudes of Rashba and Dresselhaus interactions \citep{ RevModPhys.89.011001}. This captivating phenomenon has been extensively documented in prior research \citep{ RevModPhys.89.011001,Koralek2009,Walser2012,Sasaki2014,PhysRevB.86.081306}.  The identification of equivalent strengths of the Rashba and Dresselhaus effect was experimentally confirmed  in two-dimensional electron gas semiconductor quantum-well structures, notably GaAs/AlGaAs \citep{Koralek2009,Walser2012,Sasaki2014} and InGaAs/InAlAs \citep{PhysRevB.86.081306} systems.  Observations of PST also have been extended to diverse other platforms, including the ZnO ($10\bar{1}0$) surface \citep{Absor_2015}, the strained LaAlO$_3$/SrTiO$_3$ (001) interface \citep{Yamaguchi_2017}, SnTe (001) thin films \citep{Lee_2020}, two-dimensional GaXY (X = Se, Te; Y = Cl, Br, I) systems \citep{PhysRevB.104.115145}, group - IV monochalcogenide MX monolayers (M = Sn or Ge and X = S, Se, or Te) \citep{PhysRevB.100.115104}, proustite mineral family (Ag$_3$BQ$_3$ (B = As, Sb; Q = S, Se)) \citep{PhysRevB.107.035154} $\it{etc}$.  Persistent spin helix is also realised in strain-engineered C$_{3v}$ point group materials \citep{PhysRevB.107.035155}, and also in bulk oxides (BiInO$_3$, CsBiNb$_2$O$_7$, {Hf$_{0.5}$Zr$_{0.5}$O$_{2}$}) \citep{Tao2018,PhysRevMaterials.3.084416,10.1063/5.0197098}. While non-symmorphic symmetry is expected to play an important role for the realisation of PST \citep{Tao2018}, it has been shown recently that the orbital character of the bands around the chosen k-point are also crucial  to enforce PST \citep{PhysRevB.103.014105}. Further it has been shown symmetry induced purely cubic splitting in some non-centrosymmetric systems may also host persistent spin textures \citep{PhysRevLett.125.216405}. 
\\
\indent In the present {paper}, we suggest another route to stabilize persistent spin texture in non-centrosymmetric systems. Non-centrosymmetric systems from the perspective of symmetry may be classified either as chiral or non-chiral, with polar as well as non-polar point group symmetry \citep{PhysRevB.101.014109,PhysRevB.95.245141,Sanchez2019,Fabini2024,PhysRevB.108.245146}. Chiral materials has a well-defined handedness due to the absence of inversion, mirror or other roto-inversion symmetries \citep{AdvancedMaterials.36.2308746,ma15175812}. In particular for non-polar chiral systems,  both breaking of inversion symmetry and the presence of chiral symmetries contribute to the generation of chirality-induced spin textures (STs) \citep{Ishii2014}.
\\
\indent It may be noted that, non-polar chiral systems have remained largely unexplored. In this letter, we have analysed  spin textures that can be achieved in non-polar chiral systems \cite{PhysRevB.108.L201114}. The main results of our paper may be summarized as follows. Among the available chiral point groups, we have considered the $D_{2}$ point group, where the low-energy $\bf{k.p}$ model Hamiltonian in the presence of spin-orbit coupling (SOC) has the general form  $H=E{_0}\left( k \right)+\alpha\sigma{_x}k_{x}+\beta\sigma{_y}k_{y}+\gamma\sigma{_z}k_{z}$, where $E{_0}\left(k \right)$ is the dispersion in the absence of SOC, $\alpha, \beta, \gamma $ are constants, $\sigma_{i}, k_{i}$ are Pauli spin matrices and wave vectors respectively. However depending on the orbital character of the bands belonging to a particular $1D$ representation of the $D_{2}$ point group, the Hamiltonian admits a single spin dependent term leading to PST. Interestingly at the $k_{z}=0$ plane the above Hamiltonian may be written as a linear combination of  Dresselhaus $\left(H_{D}\left(k\right)= \alpha_{D}\left(k_{x}\sigma_{x}-k_{y}\sigma_{y}\right)\right)$  and Weyl $\left(H_{W}\left(k\right)= \alpha_{W}\left(k_{x}\sigma_{x}+k_{y}\sigma_{y}\right)\right)$ Hamiltonians. It can be easily seen that when the magnitude of the strength of the Weyl ($\alpha_W$)  and Dresselhaus ($\alpha_D$) interaction parameters are comparable it is possible to stabilize PST, thereby providing an alternative route to realise PST which remained unexplored in previous studies. {Using DFT calculations we have shown that the non-symmorphic, non-polar chiral compounds Y$_{3}$TaO$_{7}$ and AsBr${_3}$ satisfies all the conditions described above to host PST around the $\Gamma$ point with $D{_2}$ symmetry for the bottom of the conduction band and top of the valence band receptively}.  Further based on symmetry arguments we have suggested possible chiral compounds that has the potential to host PST thereby enhancing the range of compounds for application in spin-orbitronics. 
%%%%%%%%%%%%%%%%%%%%%%%%%%%%%%%%%%%%%%%%%%%%%%%%%%%%%%%%%%%%%%%%%%%%%%%%%%%%%%% 
\begin{figure}[ht]
\includegraphics[width=87mm,height=87mm,keepaspectratio]{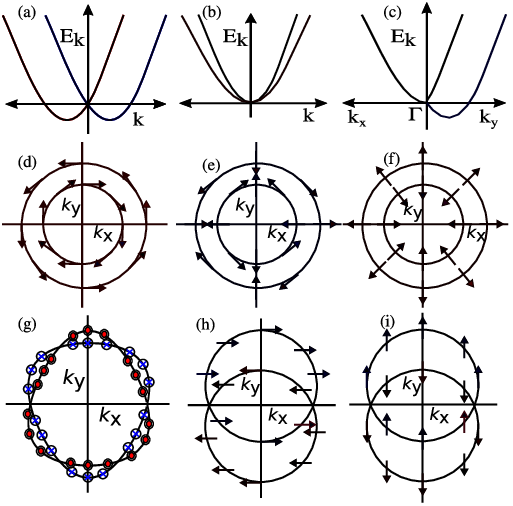}\caption{\label{fig:figure1}  Band dispersion and corresponding spin texture due to linear and cubic band splitting terms in the $k{_x}-k{_y}$ plane. (a) and (c) Illustrate the band structure ($E_{k}$ versus $k$), with $E_{k}$ = $\frac{\hbar^{2}}{2m^{\star}}k^{2}\pm\lambda_{SOC}k$ and $E_{k}$ = $\frac{\hbar^{2}}{2m^{\star}}k^{2}\pm\lambda_{SOC}k_{y}$ respectively, where $k=\sqrt{k_{x}^{2}+k_{y}^{2}}$. (b)  band dispersion due to the cubic band splitting term is $E_{k}$ = $\frac{\hbar^{2}}{2m^{\star}}k^{2}\pm\lambda_{SOC}k^{3}$. (d) Shows the schematic spin texture of the $C_{3v}$ \citep{PhysRevB.85.075404} point group for the linear Rashba model Hamiltonian , $H_{R}$, with two branches representing inner and outer branches. (e)-(f) Displays the spin texture of the $D_{2}$ point group for linear Dresselhaus $\left( H_{D} \right)$ and linear Weyl $\left( H_{W} \right)$ model Hamiltonian terms, respectively. (g) Spin texture for the  cubic model Hamiltonian $H_{C}$. Here the spin component includes solely out-of-plane (red dots) and into-the-plane (blue crosses) components. (h) Illustrates the spin texture of the $C_{2v}$ \citep{PhysRevB.85.075404} point group exhibiting PST, where the model Hamiltonian term is, $H_{PST}=\lambda_{PST}k_{y}\sigma_{x}$, realised due to comparable magnitude of  linear Rashba ($\alpha_{R}$) and linear Dresselhaus ($\alpha_{D}$) terms. (i) Shows the PST in the $D_{2}$ point group, where $H_{PST}=\lambda_{PST}k_{y}\sigma_{y}$ due to the comparable magnitude of  linear Dresselhaus ($\alpha_{D}$) and linear Weyl ($\alpha_{W}$) terms.}
\end{figure}
%%%%%%%%%%%%%%%%%%%%%%%%%%%%%%%%%%%%%%%%%%%%%%%%%%%%%%%%%%%%%%%%%%%%%%%%%%%%%%% 
\\
\indent To begin with we shall restrict  to non-polar chiral systems that will admit linear band spitting in the presence of SOC. Among the 32 crystallographic point groups only 21 point group admit linear splitting in an effective two band model Hamiltonian. Among these 21, there are a total of 6 non-polar chiral point groups such as $D_{2}$, $D_{3}$, $D_{4}$, $D_{6}$, $T$ and $O$, where linear splitting of bands may be observed \cite{PhysRevLett.125.216405,PhysRevB.108.L201114}.  
\\
\indent In Fig. $\ref{fig:figure1}$(a) and $\ref{fig:figure1}$(b), we have illustrated the band structures for linear and  purely cubic band splittings respectively for an effective two band low energy model Hamiltonian. In the case of linear band splitting, the minima of the bands are shifted. However, for cubic band splitting, $\frac{\partial E}{\partial k}=0$, at $k=0$. Fig. $\ref{fig:figure1}$(c) displays the band structure for linear spin splitting capable of hosting PST. The model Hamiltonian corresponding to linear Rashba  $\left(H_{R}\left(k\right)= \alpha_{R}\left(k_{x}\sigma_{y}-k_{y}\sigma_{x}\right)\right)$ , Dresselhaus  $\left(H_{D}\right)$, and Weyl $\left( H_{W}\right)$  give similar band splitting (see Fig. $\ref{fig:figure1}$(a)) however, the spin texture differs among them as depicted in Fig. $\ref{fig:figure1}$(d)-$\ref{fig:figure1}$(f), the two circles in the Fig. $\ref{fig:figure1}$(d)-$\ref{fig:figure1}$(f) represent two  different branches at a particular energy value. The purely cubic Hamiltonian {$H_{C}=\lambda_{C}k_{y}\left(3k_{x}^{2}-k_{y}^{2}\right)\sigma_{z}$} chosen here \citep{PhysRevLett.125.216405} exhibit spin expectation values along the z direction only, as shown in the  Fig. $\ref{fig:figure1}$(g). Both Fig. $\ref{fig:figure1}$(h) and $\ref{fig:figure1}$(i) represent the PST, with the former due to the equal and opposite strength of Dresselhaus and Rashba coupling while the latter due to equal and opposite strength of  Weyl and  Dresselhaus coupling parameter, where the  band dispersion (see Fig. $\ref{fig:figure1}$(c)) is identical for both the cases.
\\
\indent In chiral systems, the symmetry operations of the $D_{2}$ point group include 2-fold rotations along the $x \left(C_{2}^{x}\right)$, $y \left(C_{2}^{y}\right)$ , and $z \left(C_{2}^{z}\right)$ directions, as well as the identity operation. The corresponding transformation of the wave vector $\textbf{{k}}$ and Pauli spin matrices under these symmetry operations are shown in Table $\ref{tab:table1}$. The two-band model Hamiltonian {$H\left(k\right)$ for a solid} invariant under all symmetry operations listed in Table $\ref{tab:table1}$  {and unaffected by non-symmorphic \citep{PhysRevB.110.L121125} symmetry} is 
\begin{eqnarray}
H=E{_0}\left( k \right)+\alpha\sigma{_x}k_{x}+\beta\sigma{_y}k_{y}+\gamma\sigma{_z}k_{z} \label{eq:Equation1}
\end{eqnarray}.
In the $k_{x}$-$k_{y}$ plane we have,
%%%%%%%%%%%%%%%%%%%%%%%%%%%%%%%%%%%%%%%%%%%%%%%%%%%%%%%%%%%%%%%%%%%%%%%%%%%%%%%%%%%%%%%%%%% 
\begin{eqnarray}
H\left(k\right) & = & E_{0}\left(k\right)+\alpha k_{x}\sigma_{x}+\beta k_{y}\sigma_{y}\nonumber\\
 & = & E_{0}\left(k\right)+\alpha_{W}\left(k_{x}\sigma_{x}+k_{y}\sigma_{y}\right)\nonumber \\
 &  & +\alpha_{D}\left(k_{x}\sigma_{x}-k_{y}\sigma_{y}\right)\label{eq:Equation2} \\
 & =& E_{0}\left(k\right)+H_{W}\left(k\right)+H_{D}\left(k\right)\nonumber 
\end{eqnarray}
%%%%%%%%%%%%%%%%%%%%%%%%%%%%%%%%%%%%%%%%%%%%%%%%%%%%%%%%%%%%%%%%%%%
\\
\indent Here, $E_{0}\left(k\right)= E_{0}+\frac{\hbar^{2}}{2m_x^{\star}}k_x^{2}+\frac{\hbar^{2}}{2m_y^{\star}}k_y^{2}$ is the non-interacting parabolic dispersion  of the electron in the absence of SOC, with  $m_x^{\star}$ and $m_y^{\star}$ represents the effective masses of the electron, $\sigma_{x}$ and $\sigma_{y}$ are Pauli spin matrix operators. The constants  $\alpha$ and $\beta$ are given by $\alpha=\alpha_{W}+\alpha_{D}$ and $\beta=\alpha_{W}-\alpha_{D}$ where $\alpha_{W}$ and $\alpha_{D}$ are the strength of the Weyl and Dresselhaus spin orbit terms respectively. In the instance, when the strength of $\alpha_{D}$ and $\alpha_{W}$ are either equal or equal and opposite the model Hamiltonian takes the form either $H_{PST}\left(k\right)=E_0\left(k\right)+\alpha_{PST}k_x\sigma_x$ or $H_{PST}\left(k\right)=E_0\left(k\right)+\alpha_{PST}k_y\sigma_y$ respectively, where $|\alpha_W| = |\alpha_D| = \frac{\alpha_{PST}}{2}$, resulting in  PST. Diagonalizing $H_{PST}=E_{0}(k)+\alpha_{PST}k_{y}\sigma_{y}$, we obtain the eigenvalues of the model Hamiltonian, which appear as $E\left(k\right)=E_{0}\left(k\right)\pm\alpha_{PST}k_{y}$ and the dispersion for both the inner and outer branches is shown in Fig. $\ref{fig:figure1}$(c).  Our model Hamiltonian implies that the spin expectation values are $\langle \sigma_y (k) \rangle=\pm0.5$ and $\langle \sigma_x (k)  \rangle=\langle \sigma_z (k) \rangle=0.0$, which are momentum-independent, suggesting PST, as illustrated in Fig. $\ref{fig:figure1}$(i).
\\
\indent We note that  utilizing the irreducible representations (IRs) from the character table of $D_{2}$ point group (see Table $\ref{tab:table2}$  in Appendix $\ref{appendix:level1}$), we can identify the terms allowed in the low energy $\bf{k.p}$ model Hamiltonian (see Eqn. $\ref{eq:Equation1}$). The $D_{2}$ point group consists solely of 1D irreducible representations, namely $A_{1}$, $B_{1}$, $B_{2}$, and $B_{3}$. For example under $B_{2}$, both the Pauli spin component $\sigma_{y}$ and the linear momentum component $k_{y}$ remain invariant, thus allowing for the possible model Hamiltonian, to be $k_{y}\sigma_{y}$. For other irreducible representations like $B_{1}$ and $B_{3}$, alternatives such as $k_{z}\sigma_{z}$ and $k_{x}\sigma_{x}$ are only allowed respectively. The spin textures associated with such model Hamiltonian having a single spin dependent term naturally exhibit a persistent nature. Previously reported linear PSTs found under the $C_{2v}$ point group symmetry indicated that the allowed model Hamiltonian terms are $k_{x}\sigma_{y}$ and $k_{y}\sigma_{x}$ and did not include  any $k_{z}$ term. Consequently, along the $k_{z}$ direction of the BZ, linear splitting was absent \citep{Tao2018}. However, in the case of the $D_{2}$ point group, all possible linear splittings and consequent spin textures are possible. We shall argue later that the orbital character of the participating bands transforming as the basis function belonging to a particular $1D$ representation of the $D_{2}$ little group will play a crucial role in determining the terms that will be allowed in the model Hamiltonian  thereby providing a guiding principle to identify materials that may exhibit PST.
%%%%%%%%%%%%%%%%%%%%%%%%%%%%%%%%%%%%%%%%%%%%%%%%%%%%%%%%%%%%%%%%%%%%%%%%%%%%%%% 
\begin{table}[t]
\caption{\label{tab:table1} Symmetry operations of $D_{2}$ point group.}
\begin{ruledtabular}
\begin{tabular}{ccr@{\extracolsep{0pt}.}l}
\multicolumn{4}{c}{Y, $\Gamma$ point $\left(D_{2}\right)$}\tabularnewline
\hline 
$\begin{array}{c}
\text{Symmetry}\\
\text{operation}
\end{array}$ & $\left\{ k_{x},k_{y},k_{z}\right\} $ & \multicolumn{2}{c}{$\left\{ \sigma_{x},\sigma_{y},\sigma_{z}\right\} $}\tabularnewline
\hline 
$E$ & $\left\{ k_{x},k_{y},k_{z}\right\} $ & \multicolumn{2}{c}{$\left\{ \sigma_{x},\sigma_{y},\sigma_{z}\right\} $}\tabularnewline
$C_{2}^{x}$ & $\left\{ k_{x},-k_{y},-k_{z}\right\} $ & \multicolumn{2}{c}{$\left\{ \sigma_{x},-\sigma_{y},-\sigma_{z}\right\} $}\tabularnewline
$C_{2}^{y}$ & $\left\{ -k_{x},k_{y},-k_{z}\right\} $ & \multicolumn{2}{c}{$\left\{ -\sigma_{x},\sigma_{y},-\sigma_{z}\right\} $}\tabularnewline
$C_{2}^{z}$ & $\left\{ -k_{x},-k_{y},k_{z}\right\} $ & \multicolumn{2}{c}{$\left\{ -\sigma_{x},-\sigma_{y},\sigma_{z}\right\} $}\tabularnewline
\end{tabular}
\end{ruledtabular}
\end{table}
%%%%%%%%%%%%%%%%%%%%%%%%%%%%%%%%%%%%%%%%%%%%%%%%%%%%%%%%%%%%%%%%%%%%%%%%%%%%%%%
\\
\indent Based on the above general analysis, next we have identified chiral material candidates capable of hosting PST as a consequence of nearly equal and opposite strength for the Weyl and  Dresselhaus interaction parameters. Non-centrosymmetric chiral oxides are not extensively discussed in the literature for understanding the impact of chirality on the momentum-dependent band splitting. For our study, we have {first} selected the chiral oxide $\ensuremath{\text{Y}_{3}\text{Ta}\text{O}_{7}}$ as a possible candidate. $\ensuremath{\text{Y}_{3}\text{Ta}\text{O}_{7}}$ crystallizes in the base-centered orthorhombic structure  ( space group : $\ensuremath{C222_{1}}$(20)) with two formula  per unit cell. The $\ensuremath{\text{Ta}^{5+}}$ ions are bonded to six $\ensuremath{\text{O}^{2-}}$ atoms to form $\ensuremath{\text{TaO}_{6}}$ octahedra, and these octahedra exhibit chirality by sharing corners with two equivalent $\ensuremath{\text{TaO}_{6}}$ octahedra as shown in  Fig. $\ref{fig:figure3}$(a) (see Appendix $\ref{appendix:level1}$)). The BZ of the non-centrosymmetric chiral $\ensuremath{\text{Y}_{3}\text{Ta}\text{O}_{7}}$ is illustrated in Fig. $\ref{fig:figure3}$(c) (see Appendix $\ref{appendix:level1}$)), where the $\ensuremath{\Gamma}$ and Y points have $\ensuremath{D_{2}}$ little group where there is no impact of non-symmorphic symmetry see Appendix $\ref{appendix:level2}$.  The Y point is situated in the $\ensuremath{k_{y}}$-$\ensuremath{k_{z}}$ plane with $\ensuremath{k_{x}=\frac{2\pi}{a}}$. All the calculations in this {paper} are carried out within DFT with Perdew-Burke-Ernzerhof (PBE) generalised gradient approximation (GGA) \citep{PhysRevLett.77.3865} using the supplied projected augmented wave (PAW) \citep{PhysRevB.50.17953,PhysRevB.59.1758} pseudo potentials, as implemented Vienna ab initio simulation package (VASP) \citep{PhysRevB.47.558,PhysRevB.54.11169}, (for details  see Appendix $\ref{appendix:level1}$ ).
\\
\indent We have shown the non-spin polarised total DOS, orbital projected DOS and band structure of $\ensuremath{\text{Y}_{3}\text{Ta}\text{O}_{7}}$ in Fig. $\ref{fig:figure4}$ (a)-(b) of Appendix $\ref{appendix:level2}$ and find that the system is a direct band gap semiconductor with a direct band gap of 2.47 eV,  primarily centred around the Y point of the BZ. The insulating nature of the material is consistent with the nominal charge state of the system where Y is in  +3 state, Ta is in  +5 state, and O is in  -2 state. This analysis, corroborated with the partial DOS, allows us to conclude that the O-2p states are fully occupied, while the Ta-5d and Y-4d states remain  unoccupied. The TaO$_6$ octahedra are tilted toward the [001] direction, forming an infinite one-dimensional zig-zag chain. In the  octahedral environment, the Ta-$d$ states further splits into Ta-$t_{2g}$ and Ta-$e_{g}$ states, with some admixture between them in the crystal frame of reference as the z-axis of the octahedra deviates from the c-axis due to tilting. The low-lying conduction bands near the Fermi level are  primarily composed of  Ta-$t_{2g}$ states, as shown in the plot of the orbital projected band structure of Ta-$d$ states near the Fermi level in Fig. $\ref{fig:figure2}$(a).  The orbital character of the band near the $\Gamma$ point is predominantly Ta-$d_{xz}$ type while it has both Ta-$d_{xz}$ and Ta-$d_{x^2-y^2}$ character at the Y point in the crystal frame of reference. 
%%%%%%%%%%%%%%%%%%%%%%%%%%%%%%%%%%%%%%%%%%%%%%%%%%%%%%%%%%%%%%%%%%%%%%%%%%%%%%%
\begin{figure*}[ht]
\includegraphics[width=178mm,height=88mm,keepaspectratio]{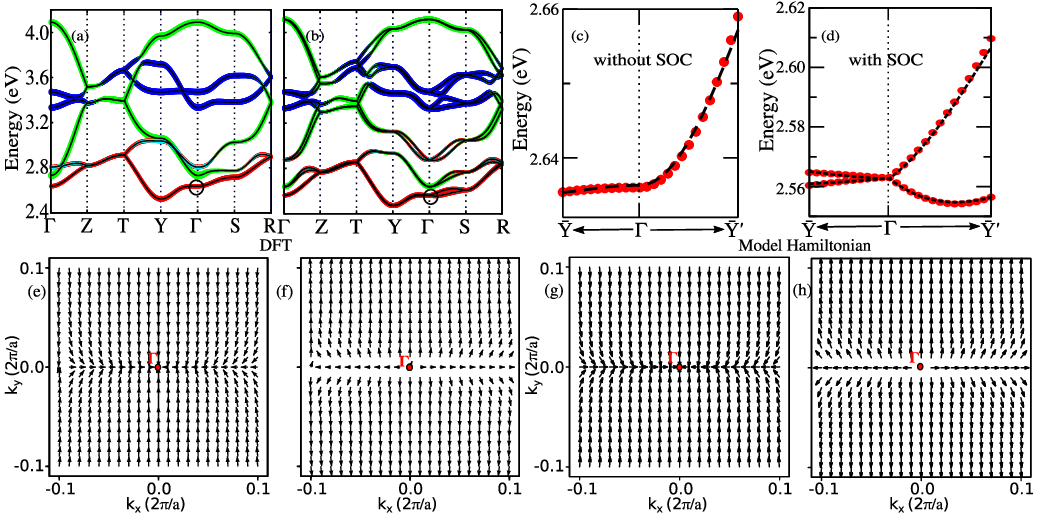}\caption{\label{fig:figure2}  (a) and (b)  Conduction bands (in black lines) with projected orbital characters, without and with SOC respectively . The Fermi levels of the systems are set to zero on the energy axis and {the bottom of the conduction band around the $\Gamma$ point is highlighted with a circle.} The orbital characters of the bands correspond to Ta-$d_{xz}$ (red), Ta-$d_{3z^2-1}$ (cyan), Ta-$d_{xy}$ (magenta), Ta-$d_{yz}$ (green), and Ta-$d_{x^2-y^2}$ (blue).(c) and (d) Illustrates band structures without and with SOC around the $\Gamma$ point {(in the bottom conduction band)} along the $\bar{Y}\left(0.15\frac{2\pi}{a},0,0\right)$-$\Gamma\left(0,0,0\right)$-$\bar{Y}^{\prime}\left(0,0.15\frac{2\pi}{b},0\right)$ direction. Circles depict the band structure obtained from  DFT, while dotted lines represent the band structure derived from the model Hamiltonian. (e) and (f) display the spin texture obtained from DFT calculations for both inner and outer branches around the $\Gamma$ {(in the bottom conduction band)} point in the $k_{x}$-$k_{y}$ plane. (g) and (h) display the same  spin texture obtained from model Hamiltonian calculations.}
\end{figure*}
%%%%%%%%%%%%%%%%%%%%%%%%%%%%%%%%%%%%%%%%%%%%%%%%%%%%%%%%%%%%%%%%%%%%%%%%%%%%%%
\\
\indent The band dispersion, total and orbital projected DOS for Y-4d, Ta-5d, and O-2p states including SOC  are presented in Fig. $\ref{fig:figure5}$ (a) and (b) of Appendix $\ref{appendix:level2}$. The introduction of  SOC results in band splitting where as expected the Ta-5$d$ states exhibit larger splitting. Fig. $\ref{fig:figure2}$(b) shows the orbital projected band structure including SOC for the Ta-$d$ states confirming that the orbital character remain same as in the calculation without SOC.  In the following we have analysed the  SOC induced band splitting and consequent spin textures around the $\Gamma$ point and  the Y point.
\\
\indent In Fig. $\text{\ref{fig:figure2}}$(c) and (d), we have shown the band structure without and including SOC for Y$_3$TaO$_7$ obtained using DFT in a narrow energy range about the $\Gamma$ point along the k-path $\bar{Y}\left(0.15\frac{2\pi}{a},0,0\right) \rightarrow \Gamma\left(0,0,0\right) \rightarrow \bar{Y}{^{\prime}}\left(0,0.15\frac{2\pi}{b},0\right)$ in the $k_{x}$-$k_{y}$ plane. It is evident that upon incorporating SOC, the bands split, with a more pronounced splitting in the $k_{y}$ direction as expected for PST (see Fig. $\ref{fig:figure1}$(c)). To gain insight into the momentum-dependent spin splitting, we have calculated the spin textures using DFT, as shown in Fig. $\text{\ref{fig:figure2}}$(e) and (f) for the inner and outer branches respectively. The spin textures appear to be nearly persistent in nature. 
\\
\indent In Fig. $\text{\ref{fig:figure2}}$(c) and (d), the dotted lines illustrate the band structure obtained from the model Hamiltonian without and with SOC respectively (see Eqn. $\ref{eq:Equation2}$). There is very good agreement between the band structure obtained from DFT and the one derived from the model Hamiltonian.  The computed values of $\alpha$ and $\beta$ are 0.02 eV $\mathring{A}$ and 0.16 eV $\mathring{A}$, respectively, resulting in the values of $\alpha_{W}$ and $\alpha_{D}$ , to be  approximately 0.09 eV $\mathring{A}$ and -0.07 eV $\mathring{A}$, respectively. The nearly equivalent values of  $\alpha_{W}$ and $\alpha_{D}$ results in PST as $H\simeq E{_0}\left( k \right)+\beta k{_y} \sigma _{y}$. It is noted that a non-zero value of $\alpha$ results in the deviation from the perfect PST in the region $k{_y}<<k_{x}$ both in the DFT calculations (see Fig. $\ref{fig:figure2}$(e) and (f)) as well as in results obtained using the model Hamiltonian (see Fig. $\ref{fig:figure2}$(g) and (h)). { Interestingly our value of the spin-orbit coupling parameter $\beta$= 160 meV $\mathring{A}$ is much larger than that reported for semiconductor well structures $\left(1-5 meV \mathring{A}\right)$} \citep{Koralek2009,Walser2012,PhysRevB.86.081306,Sasaki2014}.
\\
\indent In order  understand  the reason for the prominent contribution of  $\sigma_y k_y$ term in the model Hamiltonian, we have  analysed the orbital character of the band around the  $\Gamma$ point.  Around the $\Gamma$ point, the prominent orbital character is  Ta-$d_{xz}$, as shown in Fig. $\text{\ref{fig:figure2}}$ (a)-(b), with a small mixing of the Ta-$d_{{3z^2}-{1}}$ orbital character. The Ta-$d_{xz}$ orbital character belongs to the IR $B_{2}$ of the little group $D{_2}$. Under $B_{2}$, both $\sigma_y$ and $k_y$ remain invariant, resulting in the significant contribution of the  $\sigma_y k_y$ term in the model Hamiltonian. The small contribution of the Ta-$d_{{3z^2}-{1}}$ orbital, which belongs to the IR $A_{1}$, results in a small contribution of the  $\sigma_x k_x$ term  and is responsible for the small value of $\alpha$ and deviation from PST when $k{_y}<<k{_x}$. (see Table $\ref{tab:table2}$ in Appendix $\ref{appendix:level1}$).
\\
\indent A similar analysis for the conduction band minimum around the Y point with $D_{2}$ symmetry reveal that the orbital character of the band is Ta-d$_{xz}$ and Ta-d$_{x^2-y^2}$ where $\alpha_{W}$  is calculated to be three times larger than $\alpha{_D}$ resulting in a predominantly radial(Weyl) spin texture [see  Appendix $\ref{appendix:level2}$ for details]. 
\\
\indent  Our analysis reveal that the most general Hamiltonian (see Eqn. \ref{eq:Equation1}) invariant under $D_2$ point group symmetry has all necessary ingredients to host PST provided some additional symmetries impose further restriction and allows only one of the spin dependent term in the Hamiltonian. We have argued that the symmetry of the participating orbitals belonging to a particular $1D$ representation of the $D{_2}$ point group play a crucial role for the realisation of PST. 
\\
\indent {In order to provide further credence to our generic prescription for the realisation of PST, in addition to Y$_3$TaO$_7$, we have investigated another non-symmorphic  non-polar compound AsBr$_3$. AsBr$_3$ adopts an ammonia-like structure and crystallizes in the orthorhombic $P2{_1}2{_1}2{_1} \left(19\right)$  space group (see  Appendix $\ref{appendix:level3}$), as shown in Fig. $\ref{fig:figure7}$(a) (see Appendix $\ref{appendix:level3}$). The electronic structure of AsBr$_3$ consists of fully occupied Br-$4p$ states and completely empty As-$4p$ states consistent with the nominal charge state of the system As$^{3+}$Br$^{-}$ (for details see Appendix $\ref{appendix:level3}$)}. The {top of the} valence band around the $\Gamma$ point not only has $D_{2}$ little group but also composed of bands with predominantly Br-$p{_x}$ orbital character belonging to the $B{_3}$ IR of the $D{_2}$ point group. As a consequence  the symmetry allowed model Hamiltonian around the $\Gamma$ point is H = $\alpha \sigma_{x}k{_x}$ exhibiting PST (see  Appendix $\ref{appendix:level3}$). 
\\
\indent In conclusion, our calculation provide convincing evidence that chiral materials with $D_2$ little group  are ideal candidates for the realisation of PST. Among the chiral point groups, except for $D_3$, all other point groups have $D_2$ as a subgroup. This suggests that if a compound {lacks $D_2$ symmetry at the $\Gamma$ point in the BZ but exhibits symmetries such as $D_4$, $D_6$, $T$, or $O$, it is likely that an exploration across the BZ may reveal  $D_2$ symmetry at some other high point of the BZ. At the identified high symmetry point with $D_2$ symmetry which remain unaffected by the non-symmorphic symmetry operations and with the appropriate orbital character of the bands PST may be realised}. These new class of compounds capable of hosting PST discovered in the present work holds significant promise to display extended spin relaxation lifetime, a crucial attribute for applications in spin-orbitronic devices. 
\\
\indent K.D thanks Council of Scientific and Industrial Research (CSIR) India for support through a fellowship(File No. 09/080(1178)/2020-EMR-I). I.D. thanks Science and Engineering Research Board (SERB) India (Project No. CRG/2021/003024) and Technical Research Center, Department of Science and Technology (DST),  government of India for support. 

%%%%%%%%%%%%%%%%%%%%%%%%%%%%%%%%%%%%%%%%%%%%%%%%%%%%%%%%%%%%%%%%%%%%%
\appendix

\section{\label{appendix:level1}Structural and Computational Details}
\begin{figure}[h]
\includegraphics[width=90mm,height=76mm,keepaspectratio]{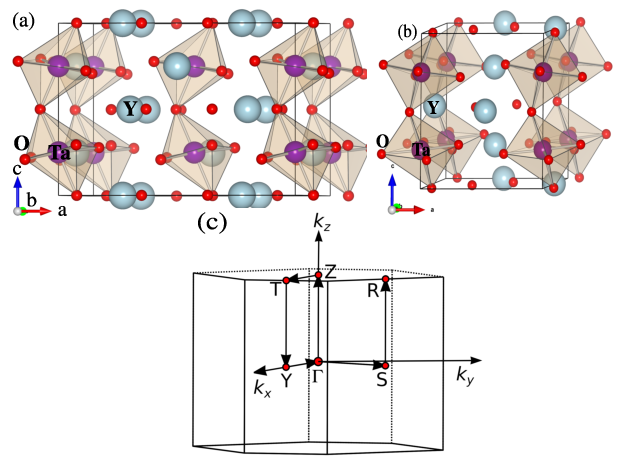}\caption{\label{fig:figure3} Crystal structure and BZ of $\text{Y}_{3}\text{Ta}\text{O}_{7}$. (a)The conventional unit cell of the bulk non-centrosymmetric chiral material $\text{Y}_{3}\text{Ta}\text{O}_{7}$. Notably, the octahedra containing the Ta atoms form a distinctive zig-zag network. (b) The primitive unit cell of the bulk non-centrosymmetric chiral material $\text{Y}_{3}\text{Ta}\text{O}_{7}$. (c) Illustrates the BZ of the base centred non-centrosymmetric chiral compound, $\text{Y}_{3}\text{Ta}\text{O}_{7}$. }
\end{figure}

Y$_3$TaO$_7$ crystallizes in a base-centered orthorhombic structure with the space group C222$_1$ (space group no. 20). The crystallographic conventional unit cell has lattice parameters $a=10.4762 $ \text{\AA}, $b=7.4237 $ \text{\AA}, and $c=7.4522 $ \text{\AA}, as summarised in Table $\text{\ref{tab:table3}}$ \citep{ALLPRESS1979105,ROSSELL1979115}. All electronic structure calculations in the present work are carried out using these lattice parameters and atomic positions. A total of 22 atoms are present in the unit cell, which corresponds to 2 formula units. The nonsymmorphic symmetry operations of the space group include a twofold rotation about the $z$-axis followed by a $c/2$ translation along the $z$-axis and a twofold rotation about the $y$-axis followed by a $c/2$ translation along the $z$-axis.
\\
\indent The crystal structure for the conventional unit cell and primitive unit cell are displayed in Fig. {\ref{fig:figure3}}(a) and Fig. {\ref{fig:figure3}}(b) respectively. Fig. {\ref{fig:figure3}}(c) represents the Brillouin zone (BZ) of the base-centered orthorhombic compound Y\textsubscript{3}TaO\textsubscript{7}. The various high-symmetry points of the BZ include $\Gamma (0.0,0.0,0.0)$ at the center of the BZ,  $Y \left(\frac{2\pi}{a},0.0,0.0\right)$ at the center of each rectangular face, $Z \left(0.0,0.0,\frac{\pi}{c}\right)$ at the center of each hexagonal face, $S$ at the center of each square face,  $T \left(\frac{2\pi}{a},0.0,\frac{\pi}{c}\right)$ at the middle of an edge joining a hexagonal face and a rectangular face, and $R \left(\frac{\pi}{a},\frac{\pi}{b},\frac{\pi}{c}\right)$ at the middle of an edge joining a hexagonal face and a square face, all of which are time-reversal (TR) invariant. Among the high-symmetry points, $Z$, $T$, and $R$ lie in the $k_z = \pi/c$ plane, while $\Gamma$, $Y$, and $S$ lie in the $k_z = 0$ plane, with the $Y$ point being 2-fold degenerate.
\\
\indent The calculations presented in this paper are performed using the Vienna ab initio simulation package (VASP) \citep{PhysRevB.47.558,PhysRevB.54.11169} within the framework of density-functional theory (DFT). We utilized the  projector-augmented-wave pseudo potentials \citep{PhysRevB.50.17953,PhysRevB.59.1758} and employed the Perdew-Burke-Ernzerhof generalized gradient approximation (GGA) \citep{PhysRevLett.77.3865}. For these calculations, the energy cut-off was set to 600 eV, and a 12\texttimes 12\texttimes 10 k-point mesh was used for the self-consistent calculations, employing the Monkhorst grid for k-point sampling.
%%%%%%%%%%%%%%%%%%%%%%%%%%%%%%%%%%%%%%%%%%%%%%%%%%%%%%%%%%%%%%%%%%%%%

%%%%%%%%%%%%%%%%%%%%%%%%%%%%%%%%%%%%%%%%%%%%%%%%%%%%%%%%%%%%%%%%%%%%%
\begin{table}[h]
\caption{\label{tab:table2}Character table for point group $D_{2}$. \cite{Dresselhaus2008}}
\begin{ruledtabular}
\begin{tabular}{c|c|c|cccc}
\multicolumn{3}{c}{$D_{2}\left(222\right)$} & $E$ & $C_{2}^{z}$ & $C_{2}^{y}$ & $C_{2}^{x}$\tabularnewline
\hline
$x^{2},y^{2},z^{2}$ &  & $A_{1}$ & $1$ & $1$ & $1$ & $1$\tabularnewline
$xy$ & $\sigma_{z},z$ & $B_{1}$ & $1$ & $1$ & $-1$ & $-1$\tabularnewline
$xz$ & $\sigma_{y},y$ & $B_{2}$ & $1$ & $-1$ & $1$ & $-1$\tabularnewline
$yz$ & $\sigma_{x},x$ & $B_{3}$ & $1$ & $-1$ & $-1$ & $1$\tabularnewline
\end{tabular}
\end{ruledtabular}
\end{table}
%%%%%%%%%%%%%%%%%%%%%%%%%%%%%%%%%%%%%%%%%%%%%%%%%%%%%%%%%%%%%%%%%%%%%

%%%%%%%%%%%%%%%%%%%%%%%%%%%%%%%%%%%%%%%%%%%%%%%%%%%%%%%%%%%%%%%%%%%%%
\begin{table}[t]
\caption{\label{tab:table3} Crystal Structure and Wyckoff Positions of the
bulk non-centrosymmetric chiral material $\text{Y}_{3}\text{Ta}\text{O}_{7}$. \citep{ALLPRESS1979105,ROSSELL1979115}}
\begin{ruledtabular}
\begin{tabular}{lccc}
\multicolumn{4}{c}{$\text{Y}_{3}\text{TaO}_{7}$}\tabularnewline
\hline
Space group & a $\mathring{\left(A\right)}$ & b $\mathring{\left(A\right)}$ & c $\mathring{\left(A\right)}$\tabularnewline
\hline
$C222_{1}$ & 10.4762 & 7.423 & 7.4522\tabularnewline
\hline
\hline
Atom & x & y & z\tabularnewline
\hline
Ta (4b) & 0.0 & 0.0 & 0.25\tabularnewline
Y I (4b) & 0.0  & 0.494 & 0.25\tabularnewline
Y II (8c) & 0.2360 & 0.2379 & 0.0\tabularnewline
O Ia (8c) & 0.143 & 0.185 & 0.278\tabularnewline
O Ib (8c) & 0.108 & 0.773 & 0.292\tabularnewline
O IIa (4c) & 0.130 & 0.5 & 0.0\tabularnewline
O IIb (4c) & 0.149 & 0.5 & 0.5\tabularnewline
O III (4c) & 0\@.0778 & 0.0 & 0.0
\end{tabular}
\end{ruledtabular}
\end{table}
%%%%%%%%%%%%%%%%%%%%%%%%%%%%%%%%%%%%%%%%%%%%%%%%%%%%%%%%%%%%%%%%%%%%%

%%%%%%%%%%%%%%%%%%%%%%%%%%%%%%%%%%%%%%%%%%%%%%%%%%%%%%%%%%%%%%%%%%%%%
%%%%%%%%%%%%%%%%%%%%%%%%%%%%%%%%%%%%%%%%%%%%%%%%%%%%%%%%%%%%%%%%%%%%%
\section{\label{appendix:level2}Electronic Structure of $\text{Y}{_3}\text{TaO}{_7}$}
\subsection{Band dispersion for Y${_3}$TaO${_7}$ without and with SOC}
\begin{figure}[h]
\includegraphics[width=8.5cm,height=8.5cm,keepaspectratio]{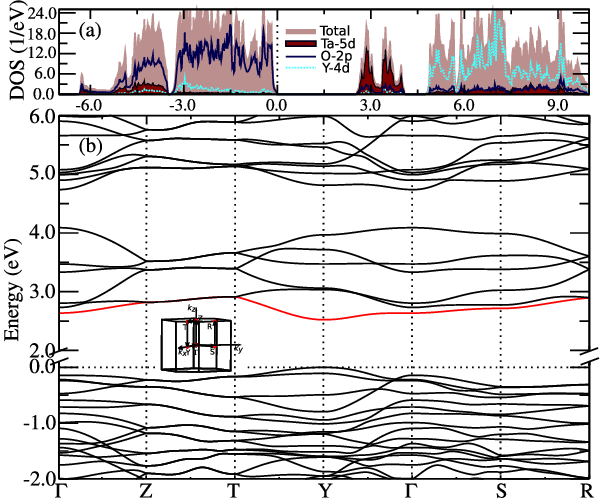}\caption{\label{fig:figure4} DOS and band structure of $\text{Y}_{3}\text{Ta}\text{O}_{7}$ without SOC. (a) Illustrates the total and partial DOS for the non-spin-polarized system $\text{Y}_{3}\text{Ta}\text{O}_{7}$. The total DOS is indicated in gray color, Ta-5d states are shown in maroon, O-2p states in blue, and Y-4d states in cyan color. (b) The band structure is plotted along the high-symmetry path $\Gamma(0.0,0.0,0.0)$-Z$\left(0.0,0.0,\frac{\pi}{c}\right)$-T$\left(\frac{2\pi}{a},0.0,\frac{\pi}{c}\right)$-Y$\left(\frac{2\pi}{a},0.0,0.0\right)$-$\Gamma(0.0,0.0,0.0)$-S$\left(\frac{\pi}{a},\frac{\pi}{b},0.0\right)$-R$\left(\frac{\pi}{a},\frac{\pi}{b},\frac{\pi}{c}\right)$. All coordinates are represented in the Cartesian lattice coordinate system. In the inset, the BZ and the path chosen to plot the band structure are shown.}
\end{figure}

\begin{figure}[h]
\includegraphics[width=8.5cm,height=8.5cm,keepaspectratio]{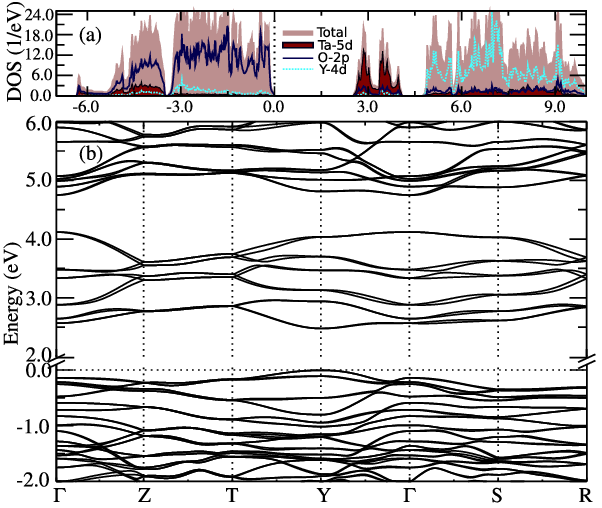}\caption{\label{fig:figure5}  DOS and band structure of $\text{Y}_{3}\text{Ta}\text{O}_{7}$ with SOC. (a) Illustrates the total and partial DOS for the non-spin-polarized system $\text{Y}_{3}\text{Ta}\text{O}_{7}$. The total DOS is indicated in gray color, Ta-5d states are shown in maroon, O-2p states in blue, and Y-4d states in cyan color. (b) The band structure is plotted along the high-symmetry path $\Gamma(0.0,0.0,0.0)$-Z$\left(0.0,0.0,\frac{\pi}{c}\right)$-T$\left(\frac{2\pi}{a},0.0,\frac{\pi}{c}\right)$-Y$\left(\frac{2\pi}{a},0.0,0.0\right)$-$\Gamma(0.0,0.0,0.0)$-S$\left(\frac{\pi}{a},\frac{\pi}{b},0.0\right)$-R$\left(\frac{\pi}{a},\frac{\pi}{b},\frac{\pi}{c}\right)$. All coordinates are represented in the Cartesian lattice coordinate system.}
\end{figure}

Fig. {\ref{fig:figure4}}(a) shows the total as well as partial density of states (DOS) and Fig. {\ref{fig:figure4}}(b) shows the band structure along the various high symmetry points of the BZ for Y${_3}$TaO${_7}$, without including spin-orbit coupling (SOC).
\\
\indent Fig. {\ref{fig:figure5}}(a) shows the total as well as partial DOS and Fig. {\ref{fig:figure5}}(b) shows the band structure along the various high symmetry points of the BZ for Y${_3}$TaO${_7}$, including SOC. A comparison of Fig. \ref{fig:figure4}(b) and \ref{fig:figure5}(b) shows that the degeneracy of the band is present even in the presence  of SOC along the path ZT in the $k{_z}=\frac{\pi}{c}$ plane due to the  nonsymmorphic symmetry $S_z = \{C_{2z}|00\frac{1}{2}\}$ \citep{Chang2018}.
However in this work, we have only analysed the band dispersion around the $\Gamma$ and Y points where there is no impact of the nonsymmorphic symmetry.
%%%%%%%%%%%%%%%%%%%%%%%%%%%%%%%%%%%%%%%%%%%%%%%%%%%%%%%%%%%%%%%%%%%%%
%%%%%%%%%%%%%%%%%%%%%%%%%%%%%%%%%%%%%%%%%%%%%%%%%%%%%%%%%%%%%%%%%%%%%
%%%%%%%%%%%%%%%%%%%%%%%%%%%%%%%%%%%%%%%%%%%%%%%%%%%%%%%%%%%%%%%%%%%%%
%%%%%%%%%%%%%%%%%%%%%%%%%%%%%%%%%%%%%%%%%%%%%%%%%%%%%%%%%%%%%%%%%%%%%
\begin{figure}[h]
\centering
\includegraphics[width=87mm,height=134 mm,keepaspectratio]{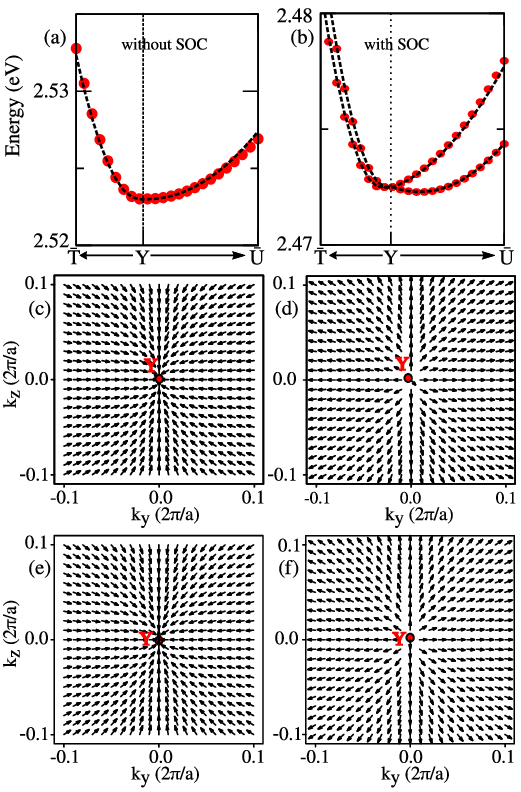}\caption{\label{fig:figure6}
Band structure in narrow range without and with SOC, along with spin textures(STs) around the Y point of Y${_3}$TaO${_7}$. (a) and (b) illustrates the band structure both without and with the inclusion of SOC effects around the Y point along the $\bar{T}\left(\frac{2\pi}{a},0,0.1\frac{\pi}{c}\right)$-Y$\left(\frac{2\pi}{a},0,0\right)$-$\bar{U}\left(\frac{2\pi}{a},0.1\frac{2\pi}{b},0\right)$ direction. Circles depict the band structure obtained from DFT, while dotted lines represent the band structure derived from the model Hamiltonian. (c) and (d) display the ST obtained from DFT calculations for both inner and outer bands around the Y point in the $k_{y}$-$k_{z}$ plane with $k_{x}=\frac{2\pi}{a}$. (e) and (f) display the ST obtained from model Hamiltonian calculations.}
\end{figure}
%%%%%%%%%%%%%%%%%%%%%%%%%%%%%%%%%%%%%%%%%%%%%%%%%%%%%%%%%%%%%%%%%%%%%
\subsection{Band dispersion and Spin texture at Y point}
In this section we have analysed the ST at the Y point.
\\
\indent Fig. \ref{fig:figure6}(a) and (b) shows the band structure within a narrow energy range around the Y point along the path $\bar{T}\left(\frac{2\pi}{a},0,0.1\frac{\pi}{c}\right)$ $\rightarrow$ Y$\left(\frac{2\pi}{a},0,0\right)$ $\rightarrow$ $\bar{U}\left(\frac{2\pi}{a},0.1\frac{2\pi}{b},0\right)$ in the $k_{x}=\frac{2\pi}{a}$ plane, in the absence and presence of SOC, respectively.  The ST of the inner and outer branches of the conduction band minimum (CBM) around the Y point obtained from DFT calculation is shown in Fig. \ref{fig:figure6}(c) and (d) in the $k_{y}$-$k_{z}$ plane respectively. The ST reveals that the spin expectation values $\langle \sigma_y  \rangle$ and $\langle \sigma_z  \rangle$ vary with momentum, while there is no $\langle \sigma_x  \rangle$ component. From Fig. \ref{fig:figure6}(c) and (d), we observe that the spins are mostly parallel to $\mathbf{k}$. For the inner band, the spins point inward, whereas for the outer band, the spins point outward. These are characteristic signatures of the linear Weyl and Dresselhaus effect \citep{PhysRevLett.114.206401,PhysRev.100.580}. To further understand the DFT results, we have derived a low-energy $\bf{k.p}$ model Hamiltonian around the Y point in the $k_{x}=\frac{2\pi}{a}$ plane,
%%%%%%%%%%%%%%%%%%%%%%%%%%%%%%%%%%%%%%%%%%%%%%%%%%%%%%%%%%%%%%%%%%%%%
\begin{eqnarray}
H & = & E_{0}\left(k\right)+\alpha_{W}\left(k_{y}\sigma_{y}+k_{z}\sigma_{z}\right)\nonumber \\
 &  & +\alpha_{D}\left(k_{z}\sigma_{z}-k_{y}\sigma_{y}\right)\label{eq:equation1} \\
 &  =& E_{0} \left(k\right) + \beta k_{y} \sigma_{y} + \gamma k_{z} \sigma_{z} \nonumber 
\end{eqnarray}
%%%%%%%%%%%%%%%%%%%%%%%%%%%%%%%%%%%%%%%%%%%%%%%%%%%%%%%%%%%%%%%%%%%%%
\\
\indent Here, $E_{0}\left(k\right)$ represents the non-interacting parabolic dispersion of the electron and along the chosen direction the parabolic dispersion is not symmetric and is written as $E_{0}\left(k\right) = E_0 + \frac{\hbar^{2}k_{y}^{2}}{2m_{y}^{\star}}+\frac{\hbar^{2}k_{z}^{2}}{2m_{z}^{\star}}$.  The band structure obtained from the model Hamiltonian is plotted by dotted lines and agrees well with the DFT band structure, plotted with circles, as shown in Fig. \ref{fig:figure6}(a) and (b). Similarly, the ST obtained from the model Hamiltonian agrees well with the ST obtained from DFT, as shown in Fig. \ref{fig:figure6}(e) and (f). The interaction parameter for the model Hamiltonian $\alpha_{W}$ and $\alpha_{D}$ are calculated to be 0.03 eV$\mathring{A}$ and 0.01 eV$\mathring{A}$ and are much smaller in comparison to the $\Gamma$ point. The values of $\beta$ and $\gamma$ are calculated to be 0.04 eV $\mathring{A}$ and 0.02 eV $\mathring{A}$ respectively. The presence of Ta-$d_{xz}$ character and $d_{x{^2}-y{^2}}$ character at the Y point admit $ k{_y} \sigma{_y}$ and $ k{_z} \sigma{_z}$ term in the low energy $\bf{k.p}$ model Hamiltonian resulting in a nearly radial (Weyl) spin texture.
\section{\label{appendix:level3} Electronic Structure and Spin texture of $\text{AsBr}_3$}
%%%%%%%%%%%%%%%%%%%%%%%%%%%%%%%%%%%%%%%%%%%%%%%%%%%%%%%%%%%%%%%%%%%%%
\begin{figure}[h]
\includegraphics[width=80mm,height=40mm,keepaspectratio]{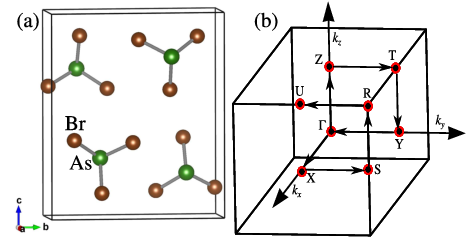}\caption{\label{fig:figure7} Crystal structure and BZ of AsBr$_3$. (a) Depicts the primitive unit cell of the bulk non-centrosymmetric chiral material AsBr$_3$. (b) Illustrates the BZ of the primitive non-centrosymmetric chiral compound, AsBr$_3$. }
\end{figure}
%%%%%%%%%%%%%%%%%%%%%%%%%%%%%%%%%%%%%%%%%%%%%%%%%%%%%%%%%%%%%%%%%%%%%
%%%%%%%%%%%%%%%%%%%%%%%%%%%%%%%%%%%%%%%%%%%%%%%%%%%%%%%%%%%%%%%%%%%%%
%%%%%%%%%%%%%%%%%%%%%%%%%%%%%%%%%%%%%%%%%%%%%%%%%%%%%%%%%%%%%%%%%%%%%
AsBr$_3$ adopts an ammonia-like structure and crystallizes in the orthorhombic $P2_12_12_1\left(19\right)$ space group \citep{AsBr3,Jain_10}, as shown in Fig. \ref{fig:figure7}(a). The structure is zero-dimensional and consists of four AsBr$_3$ clusters. As$^{3+}$ is bonded in a distorted trigonal non-coplanar geometry to three Br$^{-}$ atoms. There are a total of four formula units of AsBr$_3$ present in the primitive unit cell. The BZ of the system is depicted in Fig. \ref{fig:figure7}(b).
%%%%%%%%%%%%%%%%%%%%%%%%%%%%%%%%%%%%%%%%%%%%%%%%%%%%%%%%%%%%%%%%%%%%%
%%%%%%%%%%%%%%%%%%%%%%%%%%%%%%%%%%%%%%%%%%%%%%%%%%%%%%%%%%%%%%%%%%%%%
\begin{figure}[h]
\includegraphics[width=8.5cm,height=8.0cm,keepaspectratio]{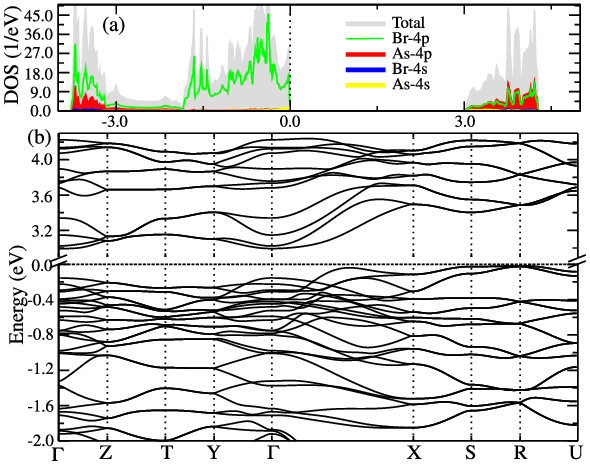}\caption{\label{fig:figure8} The  DOS and band structure of AsBr$_3$ without SOC. (a) Illustrates the non-spin-polarized total and partial  DOS. The total DOS is indicated by gray, As-4p states are shown in red, Br-4p states in green, and As-4s and Br-4s states are shown in yellow and blue, respectively. (b)  The band structure is plotted along the high-symmetry paths $\Gamma(0.0,0.0,0.0)$ - Z$\left(0.0,0.0,\frac{\pi}{c}\right)$ - T$\left(0.0,\frac{\pi}{b},\frac{\pi}{c}\right)$ - Y$\left(0.0,\frac{\pi}{b},0.0\right)$-$\Gamma(0.0,0.0,0.0)$ - X$\left(\frac{\pi}{a},0.0,0.0\right)$ - S$\left(\frac{\pi}{a},\frac{\pi}{b},0.0\right)$ - R$\left(\frac{\pi}{a},\frac{\pi}{b},\frac{\pi}{c}\right)$ - U$\left(\frac{\pi}{a},0.0,\frac{\pi}{c}\right)$. All coordinates are shown in the Cartesian lattice coordinate system.}
\end{figure}
%%%%%%%%%%%%%%%%%%%%%%%%%%%%%%%%%%%%%%%%%%%%%%%%%%%%%%%%%%%%%%%%%%%%%%%%%%%%%%%
%%%%%%%%%%%%%%%%%%%%%%%%%%%%%%%%%%%%%%%%%%%%%%%%%%%%%%%%%%%%%%%%%%%%%
\begin{figure}[t]
\includegraphics[width=8.5cm,height=8.0cm,keepaspectratio]{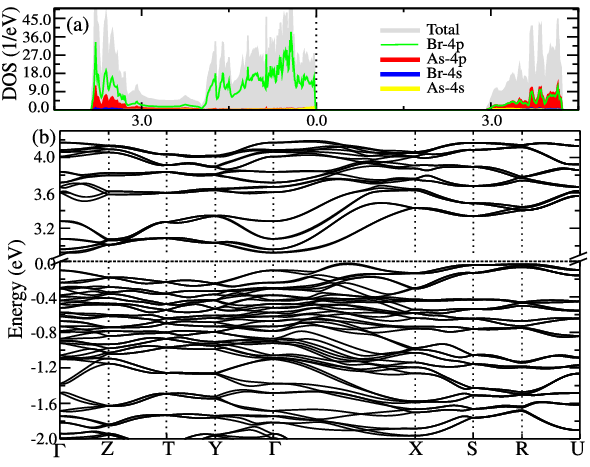}\caption{\label{fig:figure9}  The DOS and band structure of AsBr$_3$ with SOC. (a) Illustrates the spin-polarized total and partial DOS. The total DOS is indicated by gray, As-4p states are shown in red, Br-4p states in green, and As-4s and Br-4s states are represented in yellow and blue, respectively. (b)  The band structure is plotted along the high-symmetry paths $\Gamma(0.0,0.0,0.0)$-Z$\left(0.0,0.0,\frac{\pi}{c}\right)$-T$\left(0.0,\frac{\pi}{b},\frac{\pi}{c}\right)$-Y$\left(0.0,\frac{\pi}{b},0.0\right)$-$\Gamma(0.0,0.0,0.0)$-X$\left(\frac{\pi}{a},0.0,0.0\right)$-S$\left(\frac{\pi}{a},\frac{\pi}{b},0.0\right)$-R$\left(\frac{\pi}{a},\frac{\pi}{b},\frac{\pi}{c}\right)$-U$\left(\frac{\pi}{a},0.0,\frac{\pi}{c}\right)$. All coordinates are shown in the Cartesian lattice coordinate system.}
\end{figure}
%%%%%%%%%%%%%%%%%%%%%%%%%%%%%%%%%%%%%%%%%%%%%%%%%%%%%%%%%%%%%%%%%%%%%
\begin{figure}[h]
\includegraphics[width=86.5mm,height=137mm,keepaspectratio]{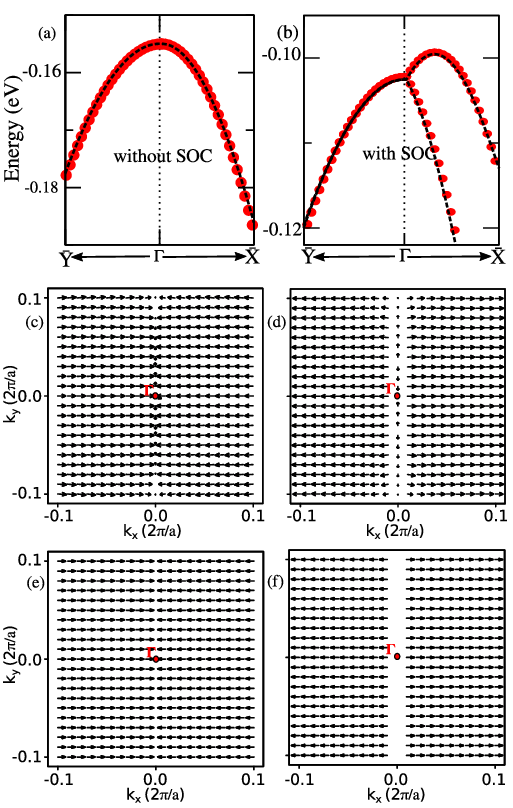}\caption{\label{fig:figure10} Band structure without and with SOC, along with STs around the $\Gamma$ point. (a) and (b) illustrates the band structure both without and with the inclusion of SOC effects around the $\Gamma$ point along the $\bar{Y}\left(0.0,0.12\frac{2\pi}{b},0\right)$-$\Gamma$ $\left(0,0,0\right)$-$\bar{X}\left(0.12\frac{2\pi}{a},0.0,0\right)$ direction. Circles depict the band structure obtained from DFT, while dotted lines represent the band structure derived from the model Hamiltonian. (c) and (d) display the ST obtained from DFT calculations for both inner and outer bands around the $\Gamma$ point in the $k_{x}$-$k_{y}$ plane. (e) and (f) display the ST obtained from model Hamiltonian calculations.}
\end{figure}
%%%%%%%%%%%%%%%%%%%%%%%%%%%%%%%%%%%%%%%%%%%%%%%%%%%%%%%%%%%%%%%%%%%%%
\\
\indent Fig. \ref{fig:figure8}(a) shows the total as well as partial DOS of the system without SOC. Here, we observe that the material is insulating in nature. This insulating behavior can be understood from the nominal charge state of the system: As has a charge state of $+3$ and Br has a charge state of $-1$. This results in fully occupied Br-4p states and completely empty As-4p states, as reflected in the DOS  (see Fig. \ref{fig:figure8}(a)). Fig. \ref{fig:figure8}(b) shows the band structure along various high-symmetry points of the BZ for AsBr$_3$, without including SOC. The band structure reveal an indirect band gap of 3.00 eV.
%%%%%%%%%%%%%%%%%%%%%%%%%%%%%%%%%%%%%%%%%%%%%%%%%%%%%%%%%%%%%%%%%%%%%
\\
\indent Fig. \ref{fig:figure9}(a) shows the total as well as partial DOS, and Fig. \ref{fig:figure9}(b) displays the band structure along various high-symmetry points of the BZ for AsBr$_3$, including SOC. As expected SOC induces splitting of the bands at various high symmetry points of the BZ. We shall focus on the valence band at the $\Gamma$ point with $D{_2}$ little group.
\\
\indent Fig. \ref{fig:figure10}(a) and (b) show the band structure within a narrow energy range around the $\Gamma$ point along the path $\bar{Y}\left(0,0.12\frac{2\pi}{b},0\right)$ $\rightarrow$ $\Gamma\left(0,0,0\right)$ $\rightarrow$ $\bar{X}\left(0.12\frac{2\pi}{a},0,0\right)$ in the $k_{x}$-$k_{y}$ plane, in the absence and presence of SOC, respectively. From the DFT band structure, we observe that the maximum is shifted from the $\Gamma$ point towards the $k_{x}$ direction, reminiscent of the PST effect. To confirm the nature of the band splitting, the ST of the inner and outer branches of the valence band around the $\Gamma$ point is shown in Fig. \ref{fig:figure10}(c) and (d) in the $k_{x}$-$k_{y}$ plane, respectively.  We observe that the spin expectation value $\langle \sigma \rangle$ is  parallel to $k_{x}$, a characteristic signature of the linear PST. To further understand the DFT results, we have derived a low-energy $\bf{k.p}$ model Hamiltonian around the $\Gamma$ point in the $k_{x}$-$k{_y}$ plane is $ H  =  E_{0}(k) + \alpha k_{x} \sigma_{x} + \beta k_{y} \sigma_{y} $.
\\
\indent The band structure obtained from the model Hamiltonian is shown by dotted lines and agrees well with the DFT band structure, shown with circles (see Fig. \ref{fig:figure10}(a) and (b)). The interaction parameters of the model Hamiltonian are calculated to be $\alpha = 0.11$ eV$\mathring{A}$ and $\beta = 0.0$ eV$\mathring{A}$ resulting in $\alpha_{W}$=$\alpha_{D}$=0.055 eV$\mathring{A}$, necessary for in PST. The ST obtained from the model Hamiltonian agrees well with the ST obtained from DFT, as shown in Fig. \ref{fig:figure10}(e) and (f).  The only contribution of the $k_x \sigma_x$ term is due to the significant contribution of the Br-$p_x$ orbital character around the $\Gamma$ point in the valence band with irreducible representation (IR) $B_3$, which only admits $k{_x} \sigma{_x}$ term in the model Hamiltonian.

%%%%%%%%%%%%%%%%%%%%%%%%%%%%%%%%%%%%%%%%%%%%%%%%%%%%%%%%%%%%%%%%%%%%%
\nocite{ALLPRESS1979105}
\nocite{ROSSELL1979115}
\nocite{Dresselhaus2008}
\nocite{PhysRevB.54.11169}
\nocite{Chang2018}
\nocite{AsBr3}
\nocite{Jain_10}

%\bibliography{30_08_2024}
\end{document}